# Quantization Mapping on Dirac Dynamics via Voltage-Driven Charge Density in Monolayer Graphene: A Klein Paradox and Entropy-Ruled Wavevector Mechanics Study

Karuppuchamy Navamani*
Department of Physics, KPR Institute of Engineering and Technology
Coimbatore-641 407, India.

*ORCID: 0000-0003-1103-2393*

*Corresponding author: pranavam5q@gmail.com

**ABSTRACT**

Thermodynamics coupled quantum features on electron and hole dynamics in Dirac materials is quite interesting and crucial for real device applications such as electronic, thermoelectric, energy devices, and quantum circuits. The correlation between the formation of electron-hole puddles near the charge neutrality point (CNP) and the role of disorder in terms of differential entropy is fundamentally important for the Dirac transport mechanism in graphene systems, but not yet well-established. With this motivation, we map the energy quantization for Dirac materials through the empirical relation between voltage-driven charge density in monolayer graphene, utilizing the differential entropy ($h_S$)-ruled wavevector ($k$) mechanics. For this work, we propose the four postulates, which are the key observable descriptions of earlier experimental and theoretical reports (appended separately), to study the precise electronic transport via an entropically guided wavevector propagation approach, along with the Klein paradox, which pertains to the ultrafast electron-hole kinetics in the Dirac or quasi-Dirac quantum systems and devices. Accordingly, the developed "$h_S$-ruled $k$" and "$h_S$-ruled $N$" relations, such as $k(h_S) = k_i exp\left(\frac{h_S}{5}\right)$ and $N = exp\left(\frac{h_S}{5}\right)$, generalize the electron dynamics in both the unbounded and potentially bounded Dirac systems, respectively. Here, $N$ is the principal quantum number, which is correlated with the wavevector. Through the quantization mapping procedure under different voltage-driven potential ($U$=$eV$) boundary conditions, the observed energy shift from lower to excited quantum state obeys the relation of $E_k = \hbar v_F k \propto N_U^3(eV_i) \rightarrow N_k(\hbar v_F k_i)$; where $N_k$ = 1, 2, 3, …, $k_i$ is the first (or initial) electronic state wavevector, and $N_U$ is the voltage-driven potential energy contribution factor for the quantum state existence, $N_{Q.St.} \equiv N_k \leftrightarrow N^3_{U=eV_g}$. In such a way, the mapped electron density, diffusion coefficient, and mobility for bounded Dirac materials are increased by orders of $N^3$, $N^2$, and $N$, respectively. This study reveals information about the interaction potential-DOS relationship in the graphene systems, which is also applicable to all other Dirac materials.

**KEYWORDS**

Differential Entropy, Wavevector, Quantization on Dirac Transport, Density of States Proportion, Voltage-Driven Charge Density, Quantum Mobility.

---

**POSTULATES**

1) The kinetic energy term is a single deterministic factor for electron (hole) dynamics, and other potential terms, such as exchange and correlations, etc., are here too negligible or cancelled each other (Ref.: Martin, J.; *et al.*, *Nature Phys.* **2008**, *4*, 144−148).

2) At a very low temperature or degenerate limit, the charge density increases in the order of $7\times10^{10}$ $cm^{-2}$ at every step of 1 Volt by applied gate-voltage (Ref.: Martin, J.; *et al.*, *Nature Phys.* **2008**, *4*, 144−148).

3) Bound-to-unbound electronic energy transition (quantization-to-quantum state collapsing) occurs by varying the boundary potential from infinite to zero or vice versa; accordingly, the charge density increases or decreases inside the well (i.e., vice versa outside the well). Here, the Klein paradox disappears in a very large supercritical potential, and the Klein paradox is recovered in a very weak or zero potential well (Refs.: Alkhateeb, M.; Matzkin, A. *Am. J. Phys.* **2022**, *90*, 297–304; Alberto, P.; *et al.*, *Eur. J. Phys.* **2018**, *39*, 025401).

4) The existence of charge density at different potentials and corresponding wavevectors $(k = \sqrt{\pi n})$ is incorporated by the differential entropy-ruled wavevector propagation relation of $k(h_S) = k_i exp\left(\frac{h_S}{5}\right)$, since the governing generalized relation for Dirac transport systems is $k(h_S) = k_i exp\left(\frac{h_S}{d+3}\right)$; where, $d$ is the dimension of the systems (Ref.: Navamani, K. *J. Phys. Chem. Lett.* **2025**, *16*, 8596−8612; K. Navamani, *J. Phys. Chem. Lett.* **2024**, *15*, 2519-2528).

---

## 1. INTRODUCTION

Over the last few decades, the two-dimensional Dirac transport materials have received much attention and great interest due to their strange electronic behaviour like ultrafast carrier dynamics (relativistic/quasi-relativistic), zero rest mass effect and quantum fluidity, etc.[1-5] The linear dispersion relationship between energy and wavevector due to massless Fermionic dynamics ($E_k = \hbar v_F k$) clearly emphasizes the non-interaction effect on electron-hole motion at the Dirac point.[1,2,6] Here, the electron-hole symmetrical transport has been principally exhibited. From one of the important research reports, it is further confirmed that the kinetic energy term acts as a key component for the transport in ultrapure monolayer graphene; and importantly, there is no other potential (electrostatic, exchange and correlation) effect involved.[6] The observation of electron hole puddles around the CNP due to disorder formation and charge density variation by external bias (electric/magnetic/doping) raises curiosity about how the interactive potential and disorder are affecting the linear dispersion (i.e., $E_k^* = \hbar v_F k^*$) as well as the Dirac transport mechanism via the wavevector propagation.[6] Accordingly, the changes of momentum (or wavevector) with respect to the external interaction directly influence the energy dispersion without affecting the linear relation, which is related to the slope of dispersion, and it leads to energy levels shift/variation. This gives information on the dynamics of electron and hole carriers in the order of Fermi velocity ($1\times10^6$ m/s), instead of carrier effective mass. For clean graphene, there is no particle at the Dirac point due to zero or vanishing density of states, which reveals the full transmission property, and it obeys the Klein paradox or tunnelling.[6,7] At this zero-boundary DP juncture (where the Fermi cone meets), the particle behaves photon-like wave packet. The potential well has been formed while applying the gate-voltage (from zero to a large/supercritical range), and hence it is expected that the presence of charge within a potential bound (like a particle in a box), at which the Klein paradox slowly disappeared.[7] The applied potential difference between an infinitesimal distance in quantum systems/devices acts as a model of a particle in a box (boundary value problem), in which the electrons are occupied in the existing quantum states. Here, the existing electrons and holes are in equal probability in the voltage range $+V$ to $-V$, or vice versa.[6-8] The formed electron-hole puddles in or nearer to the CNP have been estimated charge density with equal weightage. Based on voltage-driven charge density, the wavevector propagation has been anticipated, $k = \sqrt{\pi n}$. By introducing the boundary condition, the quantized energy for a massless (or zero rest mass) relativistic fermionic particle was solved from the Klein-Gordon equation, $E_N = NE_i = N\hbar v_F k_i$; where $N$ is the principal quantum number.[7,8] Here, the presence of electron probability density and wave packet dynamics at different voltage-driven boundaries is incorporated by the statistical and thermodynamic via the parameter differential entropy ($h_S$).[9-12] By imposing an entropy-ruled wavevector hypothesis,[9,13] the eigenvalue variation (energy shift) with potential (i.e., under boundary) is connected with the principal quantum number ($N$), $E = N\hbar v_F k_i \rightarrow \hbar v_F k_i exp\left(\frac{h_S}{5}\right)$; where, $h_S = ln\left(\sqrt{2\pi \exp(1)}\right)\sigma_{GW}$. Now, the existence of quantum states is incorporated by differential entropy ($h_S$), and hence, the total $h_S$ is the summation of all states' entropy. In this study, $\sigma_{GW}$ is the Gaussian width of the carrier wave packet, which can vary state to state. The applied potential difference between a finite or infinitesimal distance in quantum systems/devices acts as a model of a particle in a box (boundary value problem), in which the electrons are occupied in the existing quantum states. According to the potential value ($U=eV$, $e$ is an elementary electric charge) and length of the boundaries (or channel width), the $\sigma_{GW}$ will be varied, and hence the differential entropy can be measured by $h_S = ln\left(\sqrt{2\pi \exp(1)}\right)\sigma_{GW}$.[6,13] In other words, depending upon the energy state, the magnitude of $\sigma_{GW}$ is generally expected, which reveals the localization or delocalization of the carrier via the distribution over the electronic state. In this context, the bias-driven degeneracy effect on carrier dynamics can be explained by the differential entropy (see Figure 1).

## 2. CONCEPTUALIZATION AND IMPLEMENTATION

As described by previous studies,[9,10,13,14] the charge variation is exponentially compensated by differential entropy at different energetic disorder regimes; and hence it gives both the position ($x$)-momentum state-wise ($p = \hbar k$) particle information, which is correlated with the uncertainty relation of $\sigma_{GW,x}.\sigma_{GW,k} \geq \frac{\hbar}{2}$. The potential energy as a function of the position (or localized information) and kinetic energy (delocalized property) terms is associated with the wavevector ($k$). In the present study, the particle (i.e., electron) probability density inside or outside the potential well is incorporated by position or momentum dependent width of the Gaussian envelope ($\sigma_{GW,x}$ or $\sigma_{GW,k}$), and is generally reliant on a given boundary value.[6,15] In principle, the Gaussian width of an electron wave packet as a function of position or momentum is explicitly written as, $\sigma_{GW,x} \sim \frac{1}{\sigma_{GW,k}}$.[16] According to the entropically varying charge density relation, the generalized wavevector (momentum) equation for Dirac materials is described as,[9,13]

$$k(h_S) = k_i exp\left(\frac{h_S}{d+3}\right) \rightarrow N \qquad (1)$$

To this extent, the dimension-dependent $h_S$ value for Dirac quantum systems (in the general case) can be expressed as,[9]

$$h_S = (d+3) lnN \rightarrow \sigma_{GW} \equiv \frac{1}{\sqrt{2\pi \exp(1)}} N^{(d+3)} \qquad (2)$$

Therefore, the ratio between the Gaussian wave packets' width as a function of momentum, as well as position under the electronic transition ($i \rightarrow f$) situations can be mapped by

$$\frac{\sigma(p_f = \hbar k_f)_{GW}}{\sigma(p_i = \hbar k_i)_{GW}} \leftrightarrow \frac{\sigma(x_i)_{GW}}{\sigma(x_f)_{GW}} = \left(\frac{N_f}{N_i}\right)^{d+3} \qquad (3)$$

where $d$ and $N$ are the dimension (1D, 2D and 3D) and principal quantum number of a given system, respectively. Due to the effects of interaction potential and thermodynamics, the size and shape of the Gaussian width will be varied; apparently, the dynamics of charge carriers are modified, which is analyzed in Figure 1. Here, the differential entropy and chemical potential (or Fermi energy) play a vital role in describing the quantization of energy levels in the form of wavevector propagation in each eigenstate.[8,17-21] In this regard, the differential entropy-ruled wavevector for two-dimensional Dirac transport systems of molecules and materials is $k(h_{eff}) = k_i exp\left(\frac{h}{5}\right)$, and therefore the governing energy ($E = N\hbar v_F k_i$) equation is

$$E_{i\rightarrow j}(h_{S,rel}) = E_i exp\left(\frac{h_{S,rel}}{5}\right) \qquad (4)$$

Here, the $E_{i\rightarrow j}$ is the particle energy variation due to electronic transition within a bounded quantum system, and $h_{S,rel}$ is the relative differential entropy.

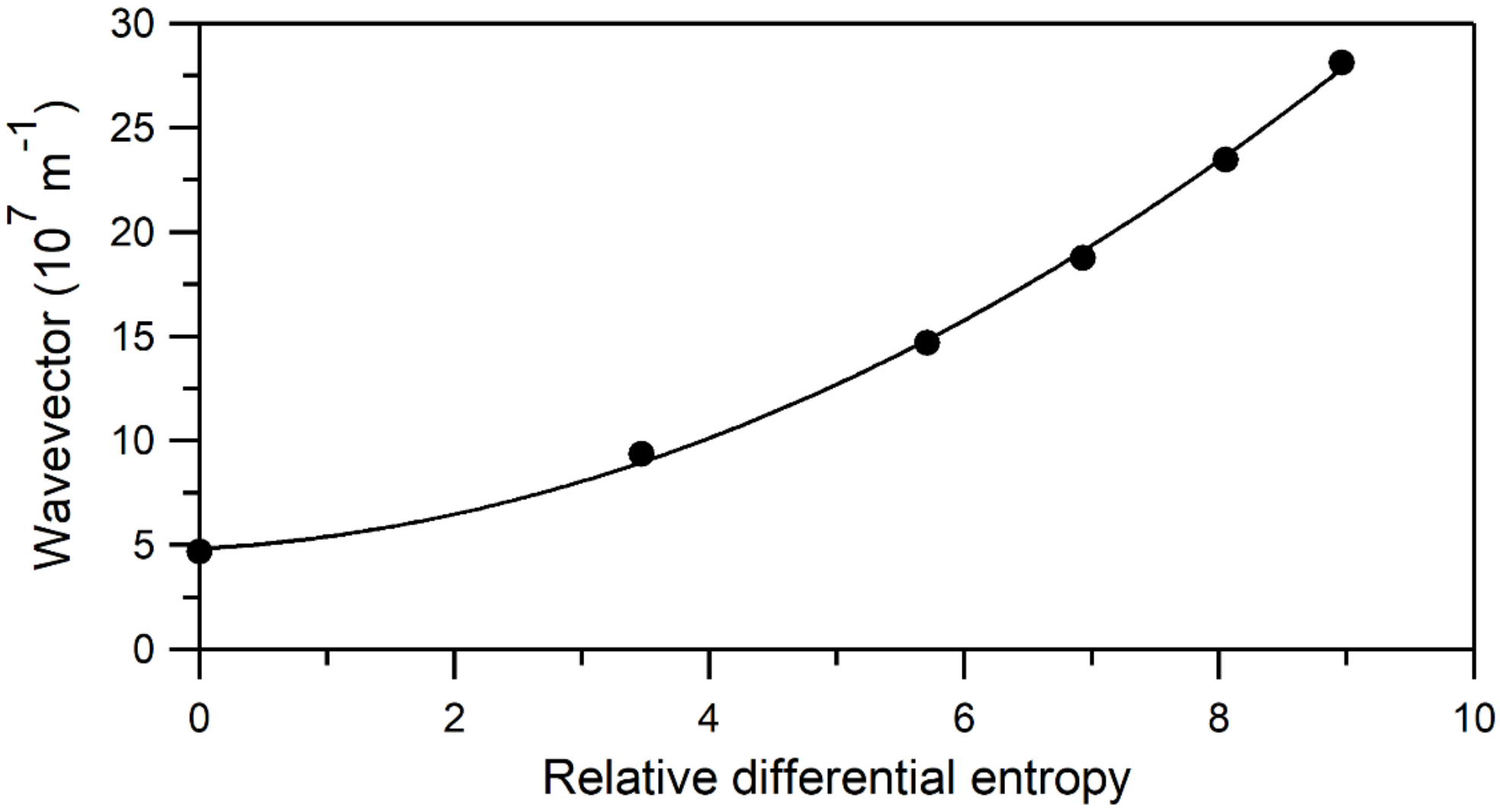

**Figure 1.** The wavevector variation with respect to the differential entropy directly gives the bias-driven degeneracy effect on electron dynamics in monolayer graphene at $T$=0.4 K. Here, the energy has a linear relationship with the wavevector (or momentum). The density of states proportion (DOSP) is principally correlated through the energy (in the form of wavevector, $E(h_S) = \hbar v_F k(h_S)$) scaled differential entropy factor, $\frac{dh_S}{dE} = (\hbar v_F)^{-1}\frac{dh_S}{dk}$. The slope of this plot here describes the DOSP, and hence the conventional DOS (or compressibility) has been mapped by the equation of $n(h_S)\left(\frac{3}{5}\right)\frac{dh_S}{dE}$ for any two-dimensional Dirac systems/devices (see Table 1 and refs. 9 and 13). The exponential variation on DOSP is directly correlated to the diffusion-based mobility for Dirac quantum systems.

Due to the carrier energy flux correlated entropy production principle, one can interpret the externally influenced electronic energy transition with entropy changes, $h_S \sim 2ln\left(\frac{dE_{i\rightarrow j}}{dE_i}\right) \rightarrow 2ln\left(\frac{E^*}{E^0}\right)$; where, $E^*$ and $E^0$ are the excited and ground state energy, respectively.[9] In this study, voltage-driven energy shifts via degeneracy strength and relevant charge density variation are estimated by the entropy-ruled method. In this connection, the density of states proportion (DOSP) is defined as the change in differential entropy per unit energy (in terms of wavevector); or simply it is described as energy-scaled entropy $\left(\frac{dh_S}{dE} \xrightarrow{Dirac} \frac{dh_S}{dk}\right)$.[9] In principle, the DOSP times charge density is the conventional DOS. Hitherto, one-to-one variation between the differential entropy and energy or chemical potential illustrates the interaction potential impact on electronic transport behaviour in any Fermionic systems of molecules and materials. As reported earlier studies,[6,9] the change of momentum (or energy; since here the linear dispersion relation of *E*-*k*) with respect to the differential entropy gives an inverse of density of states (DOS) proportion $\left(\frac{dE_k}{dh_S} \leftrightarrow \frac{dk}{dh_S}\right)$, which contains the information about the interaction potential (including external bias) effect on electron hole carrier' kinetics (see Figure 1). The DOS is an equivalent form of electronic compressibility.[6,9,22] Through compressibility or DOSP, the compressible continuum states (band) or incompressible separated states have been characterized that can hereby modify with the voltage-driven potential energy (see Figures 2 and 3). Using DOSP, the typical transport will be classified for a given electronic system/device as obeying either localized hopping or delocalized band or intermediate between them. It is important to note that the DOSP is a fundamental key descriptor for all transport quantities like mobility, conductivity, current density and diode principle, etc., since the unified version of mobility for degenerate quantum systems is $\mu = \left(\frac{d}{d+2}\right)\frac{dh_S}{dE_F}qD$.[9,13] Here, $q$, $D$, and $\frac{dh_S}{dE_F}$ are the electric charge, diffusion coefficient and DOSP, respectively. This $\frac{dh_S}{dE_F}$ factor (i.e., DOSP) carries net interactive potential (strongly or weakly correlated), bias-driven degeneracy, along with the disorder and other thermodynamics effects on electronic property, which incorporates quantum-classical transport transition too.

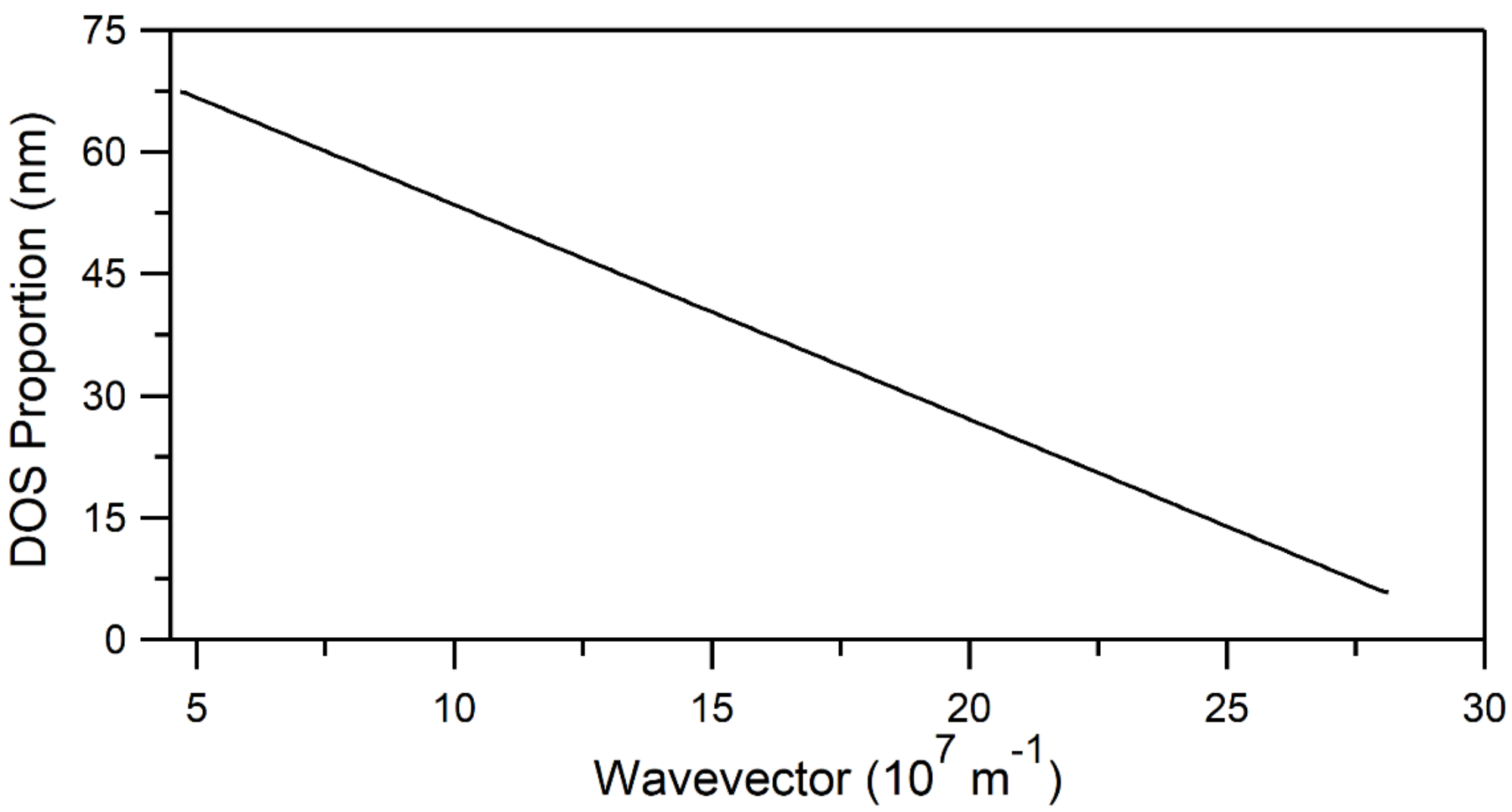

**Figure 2.** The DOSP factor $\left(\frac{dh_S}{dk}\right)$ linearly varies with the differential entropy-ruled wavevector, which is fundamentally acting as a key descriptor for diffusion-based mobility relation (or D/μ ratio) and other extended transport quantities like conductivity, current density and diode principle, etc. The changes of normal DOS due to external effects (e.g., voltage, electric field, doping and thermodynamic effect) can be quantified by this DOSP factor, which mainly contributes to electronic transport properties.

A unified version of the entropy-ruled charge transport has been parameterized through the quantities of charge density, wavevector, DOS, and quantization energy mapping for nonrelativistic dynamics materials (Schrödinger-type particle transport systems, $E \propto k^2$) and relativistic dynamics of massless Fermionic transport systems (Dirac materials, $E \propto k$) are given in Table 1.

**Table 1.** Derived generalized versions of expressions for entropy-ruled charge transport quantities for Schrödinger-type particles' transport and Dirac transport systems. In this study, the system's entropy exponentially modifies the CT quantities, which mainly accounts for the external bias-associated electronic degeneracy effect on charge transport (refs. 9, 23 and cross references therein).

| **Charge transport parameters** | **Unified formalism** | |
|---|---|---|
| | Schrödinger-type particles' transport systems (non-relativistic cases) | Dirac particles' (massless Fermionic) transport systems (relativistic cases) |
| Charge density | $\left.\frac{n(h_S)}{n_i}\right\vert_{\text{Schrödinger}} = exp\left(\frac{d}{d+2}h_S\right)$ | $\left.\frac{n(h_S)}{n_i}\right\vert_{Dirac} = exp\left(\frac{d+1}{d+3}h_S\right)$ |
| Wavevector | $\left.\frac{k(h_S)}{k_i}\right\vert_{\text{Schrödinger}} = exp\left(\frac{h_S}{d+2}\right)$<br>Here, $k(h_S) = \sqrt{2\pi n_i} exp\left(\frac{h_S}{d+2}\right)$ | $\left.\frac{k(h_S)}{k_i}\right\vert_{Dirac} = exp\left(\frac{h_S}{d+3}\right)$<br>Here, $k(h_S) = \sqrt{2\pi n_i} exp\left(\frac{h_S}{d+3}\right)$ |
| DOSP | $\frac{dh_S}{dE_F} \rightarrow \frac{m^*}{\hbar^2 k}\frac{dh_S}{dk}$ | $\frac{dh_S}{dE_F} \rightarrow \frac{1}{\hbar v_F}\frac{dh_S}{dk}$ |
| DOS (i.e., electronic compressibility) | $n(h_S)_{\text{Schrödinger}}\left(\frac{d}{d+2}\right)\frac{dh_S}{dE_F}$ | $n(h_S)_{Dirac}\left(\frac{d+1}{d+3}\right)\frac{dh_S}{dE_F}$ |
| By applying Klein paradox/ tunnelling (full transmission under unbounded situations) | -- | $\left.\frac{k(h_S)}{k_i}\right\vert_{Dirac} = exp\left(\frac{h_S}{d+3}\right)$ |
| Energy Quantization (By imposing boundary condition; i.e., absence of Klein paradox) | $E_{\text{Schrödinger}}: N^2 \propto exp\left(\frac{2h_S}{d+2}\right)$;<br>$\because E_N = \frac{N^2 h^2}{8mL^2} \rightarrow N^2 E_i$<br>where, $N$, $m$ and $L$ are the principal quantum number, particle (electron) mass and length of the potential boundary. | $E_{Dirac}: N \propto exp\left(\frac{h_S}{d+3}\right)$<br>Here,<br>$E(N) = N\hbar v_F k_i$<br>$NE_i \propto exp\left(\frac{h_S}{d+3}\right) \rightarrow N$<br>$\left.\frac{k(N)}{k_i}\right\vert_{Dirac} = N$<br>$= exp\left(\frac{h_S}{d+3}\right)$<br>$\therefore\ N_f = N_i\, exp\left(\frac{h_{S,f} - h_{S,i}}{d+3}\right)$ |

## 3. QUANTIZATION MAPPING AND ANALYSIS

The bounded and unbounded free electron dynamics in 2D Dirac materials are investigated through the differential entropy-ruled wavevector mechanics model and the Klein paradox approach. For the systems of bounded electrons and free electrons, the transport here is mapped with the equations of $k_f(h_S) = k_i exp\left(\frac{h_{S,f}-h_{S,i}}{d+3}\right)$ and $N_f = N_i\ exp\left(\frac{h_{S,f}-h_{S,i}}{d+3}\right)$, respectively, which is originally derived from the entropically varying charge density equation.[9,13] In principle, the potentially bounded Fermionic systems normally have a charge density (within a boundary), which violates the Klein paradox in which particle distribution over the existing quantized momentum states is mapped by the quantum states, $N_f = N_i\ exp\left(\frac{h_{S,f}-h_{S,i}}{d+3}\right)$ and is noted from Figure 3. On the other hand, the Klein paradox is a valid one to explore an unbounded massless free electron kinetics, and hence the vanishing or zero charge density (or DOS) is expected at this regime due to zero boundary (i.e., transmission probability → 1), which has been noted at CNP in the Dirac materials.[6,7] In this case, the electron motion is characterized by the width of the Gaussian wave packet and is incorporated by the differential entropy-ruled wavevector mechanics, $k_f(h_S) = k_i exp\left(\frac{h_{S,f}-h_{S,i}}{d+3}\right)$. Under potential well, the electronic property is mainly associated with the energy modes of the quantum states, and here the energy level separations depend upon the magnitude of the potential barrier height. It is noteworthy that, using the equation of entropy-ruled wavevector propagation, the dynamical behaviour of both free electrons and bound electrons is characterized (see Figures 1, 3 and 4), which unifies the Klein paradox as well as the potentially bounded value problems for the relativistic dynamics of massless Fermions.

Using the proposed postulates for this work, the charge dynamics parameters such as charge density, wavevector, carrier energy and differential entropy are calculated at every unit change of applied gate-voltage (i.e., $V_g$ = 1, 2, 3, …, $N$ volts), by appropriate equations from Table 1, and the results are summarized in Table 2. Through these parameters, the DOSP, quantization mapping at different voltage-driven potential, excitonic property, and potential energy contribution on energy quantization (i.e., populated states) are analysed, which are fundamental key descriptors for Dirac transport materials. As discussed above, the DOSP of any Dirac systems can be obtained by the wavevector scaled entropy $\left(\frac{dh_S}{dk}\right)$ or change in entropy per unit energy change (i.e., $\Delta E = \hbar v_F \Delta k = 1\ eV$; also see ref. 9), which is shown in Figures 1 and 2. In such that, the Dirac DOSP here is redefined as the entropy changes at each step interval of 1.5193×10$^{7}$ cm$^{-1}$ wavevector; here, $\Delta k = \frac{1}{\hbar v_F}$ cm$^{-1}$, here $\hbar v_F$ =6.582×10$^{-10}$ eV.m. The electrostatic potential and kinetic energy can be tuned by an external bias voltage and are calculated by the relation,[15]

$$U_e = eV_g - E_F(k) \rightarrow U_e = eV_g - E_k \qquad (5)$$

By implementing our postulates (postulate **1**) in the above relation (Eqn. 5), the voltage-driven potential and kinetic energy are incorporated in the following expressions as (see Table 1),

For 2D Dirac cases: $$eV_g = E_k = \hbar v_F k(h_S) \equiv \hbar v_F \sqrt{\pi n(h_S)} \rightarrow \hbar v_F \sqrt{\pi n_i} exp\left(\frac{h_S}{5}\right) \qquad (6)$$

For 2D non-Dirac cases: $$eV_g = E_k = \frac{\left(\hbar k(h_S)\right)^2}{2m} \equiv \frac{\hbar^2}{2m} 2\pi n(h_S) \rightarrow \frac{\hbar^2}{2m} 2\pi n_i \ exp\left(\frac{h_S}{2}\right) \qquad (7)$$

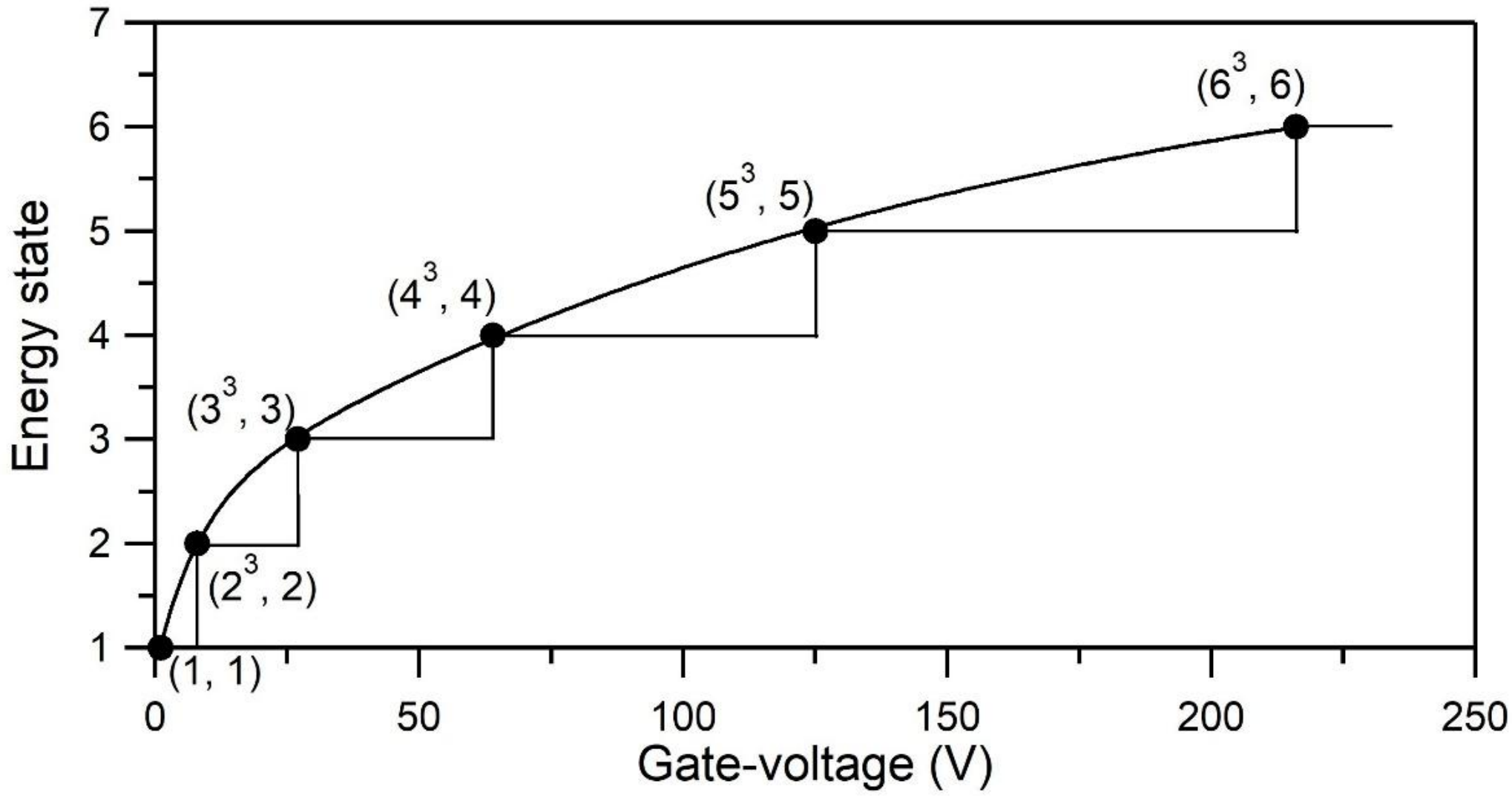

**Figure 3.** The electronic energy of electrons over the quantum states under different gate voltages ($V_g$) is mapped for potentially bounded Dirac systems. The excitonic behaviour can be illustrated by this voltage-driven potential energy ($eV_g$)-quantization energy levels ($E_N$). In monolayer graphene, the existence of quantization obeys the rule of $N_{Q.St.} \leftrightarrow N^3_{U=eV_g}$. The main described observation is that the required potential well for the quantum states' existence is about the fitting equation, $E_N \rightarrow NE_i \propto N^3(eV_i) \leftrightarrow U(V_g)$. In such a bounded electronic system of relativistic massless Fermionic systems, the Klein paradox is invalid, or the breakdown happens. The formation of gate-voltage-driven electronic levels follows a step function, with a trend similar to that in Fig. 8.36 of ref. 24.

Noteworthy, as observed by an earlier experiment which was carried out by Martin *et al.*,[6] the kinetic energy term is only involved in the ultrapure monolayer graphene system or any other ideal Dirac materials. In this context, the applied voltage directly influences the carrier dynamics via the exponential weightage of differential entropy. The existence of degeneracy modifies the Gaussian wave packet dynamics of an electron or hole, which is principally connected with the differential entropy.[9,13] With respect to the external bias, like voltage or electron/hole-rich-doping effect, the transition/shift energy scaled entropy (i.e., entropy changes per unit energy variation or shift) provides information about a newly populated or depopulated (compressible or incompressible) quantum states via DOSP.[9,23] The charge density times of DOSP, along with the dimensional factor, naturally give the real DOS of a given electronic system/devices (see Table 1).

**Table 2.** Using differential entropy-ruled charge transport and wavevector mechanism, the possible electronic states energy, wavevector, relative differential entropy, and charge density are calculated under different voltages for potentially bounded monolayer systems at 0.4 K temperature. For doubling the carrier energy (or, for *N*-multiply times carrier energy), the required voltage is $2^3$=8 V (or, $N^3$-multiply times of applied voltage will be required). In boundary cases, the existing electronic states follow the relation, $E_N(k) = NE_i \propto N^3 eV_i$. The applied voltage contribution on the inverse of DOSP here is described as $\frac{dE_N}{dh_S} = E_i \frac{dN}{dh_S} = 3eV_i N^2 \frac{dN}{dh_S} \rightarrow \hbar v_F \frac{k(h_S)}{dh_S} = \hbar v_F k_i \frac{dN}{dh_S}$; since, $N = exp\left(\frac{h_S}{5}\right)$. For ideal monolayer graphene obeys Dirac dynamics, and hence massless relativistic Fermions typically take the motion-like particle in (2d+1)-dimensional systems, due to the emergence of an additional dimension by relativistic effect (also refer, Table 1).

| Energy State (*N*) | Carrier Energy (meV) | Wavevector ($10^7$ m$^{-1}$) | Relative Differential Entropy | Charge Density ($10^{10}$ cm$^{-2}$) | Threshold Voltage (V) |
|---|---|---|---|---|---|
| 1 | 30.868 (=$E_1$) | 4.689 (=$k_1$) | 0 | 7 (=$1^3$×7) | 1 (=$1^3$) |
| 2 | 61.736 (=2$E_1$) | 9.379 (=2$k_1$) | 3.466 | 56 (=$2^3$×7) | 8 (=$2^3$) |
| 3 | 92.604 (=3$E_1$) | 14.693 (=3$k_1$) | 5.711 | 215.403 (=$3^3$×7) | 27 (=$3^3$) |
| 4 | 123.472 (=4$E_1$) | 18.759 (=4$k_1$) | 6.932 | 448.10 (=$4^3$×7) | 64 (=$4^3$) |
| 5 | 154.434 (=5$E_1$) | 23.463 (=5$k_1$) | 8.051 | 877.00 (=$5^3$×7) | 125 (=$5^3$) |
| 6 | 185.208 (=6$E_1$) | 28.134 (=6$k_1$) | 8.959 | 1512.20 (=$6^3$×7) | 216 (=$6^3$) |

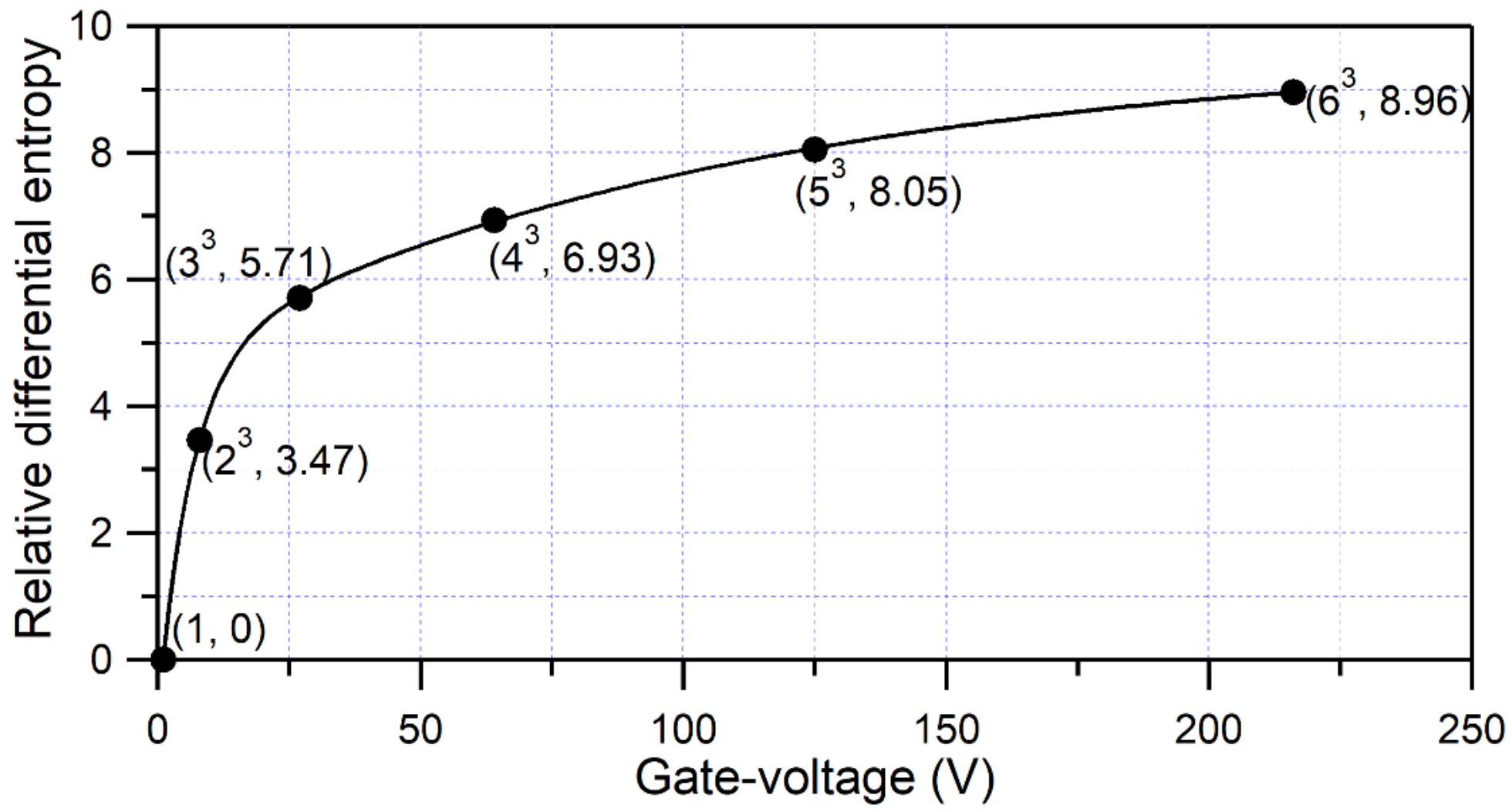

**Figure 4.** The bias-voltage effect on differential entropy, $h_S = ln\left(\sqrt{2\pi \exp(1)}\,\sigma_{GW}\right)$, provides electron and hole carriers drift, which is associated with the field-response transport property. The localization and delocalization nature of charge is generally characterized by the Gaussian width, $\sigma_{GW}$. Through the plot of gate-voltage-differential entropy relation, the compressible or incompressible electronic states can be analysed in monolayer graphene with degenerate situations ($E_f >> k_B T$). The potential energy via bias voltage contribution to the DOSP is described here as $DOSP(V_g) \rightarrow \frac{dh_S}{eV_g}$, which tells the populated electron-hole puddles at around the Dirac point.

For an unbounded case, the electron dynamics in the Dirac materials takes the propagation in the form of a Gaussian wave packet and is described hereby the equation of $k(h_S) = k_i exp\left(\frac{h_S}{5}\right)$, at which the Klein paradox is valid. The full transmission probability (i.e., $P_T \rightarrow 1$) is anticipated due to zero potential boundary or free-particle dynamics (see, ref. 7). In such a case, the Gaussian wave packet of electrons takes a continuous propagation with time (i.e., without time delay), which can be comparable with the principle of ideal delocalization in an out of box. In this study, the applied voltage-driven wavevector and related transport enhancement have an exponential function on differential entropy, which is plotted in Figure 1. That means, we principally can say that the reduction of barrier height to achieve the complete transmission of a particle processed by this "$h_S$-ruled $k$" mechanics. On the other hand, the existence of quantum states and relevant excitonic transition dynamics in the potential well is hereby illustrated by $N = exp\left(\frac{h_S}{5}\right) \rightarrow \frac{N_f}{N_i} = exp\left(\frac{h_{S,f} - h_{S,i}}{5}\right)$; where, $h_{S,f}$ represents the relatively more delocalization (i.e., more kinetic and less potential) character in the higher energy states rather than that of $h_{S,i}$ (less kinetic and more potential). Here, the dynamics of the energy packet (at each mode of momenta state) within a boundary is mapped using the above differential entropically determined energy state equation, which is the "$h_S$-ruled *N*" approach, which is shown in Figures 3 and 4. This result is in corroboration with the earlier study, which was carried out by Varlamov *et al.*[18] In this quantization condition, the Klein paradox is invalid and hence disappears. The particle probability inside (outside) of the box also revealed this $h_S$-ruled *N* ($h_S$-ruled *k)* relation with respect to the variation in Gaussian width of the wave packet, $dh_S = \frac{d\sigma_{GW}}{\sigma_{GW}} \rightarrow \frac{\sigma'_{GW}-\sigma_{GW}}{\sigma_{GW}}$. It is noted that, under different potential boundaries (1, 2, 3, 4, 5 *eV*), the existing quantization of ground state energy levels are around the values of 30.87, 38.88, 92.60, 123.47, and 154.43 meV, respectively, and other higher energy state values are follows the relation of $E(N) = N\hbar v_F k_i$; since, $N = exp\left(\frac{h_S}{5}\right)$, see and compare Figures 3 and 4. The main observation is that the formation of energy quantization with respect to the potential well for monolayer graphene (Dirac systems) is mapped by the relation of $E_N \rightarrow NE_g \propto N^3(eV_i)$; where, $E_g$ and $eV_i$ are the ground state energy and the supplied initial potential energy $(U_i)$ boundary (also refer to Table 1 and Figure 3). Here, the voltage-driven energy profile principally obeys the step function (compare with Figure 8.36 in ref. 24).[24] Besides that, the charge density (or probability density) variation at each energy state also follows the relation $E_N \rightarrow NE_g \propto N^3 n_i$, since the wavevector and charge density are exponentially increased with the $\frac{h_S}{5}$ and $\frac{3h_S}{5}$ factor $\left(\frac{k_f}{k_i} \equiv \left(\frac{n_f}{n_i}\right)^{\frac{1}{3}} = exp\left(\frac{h_{S,rel}}{5}\right)\right)$, respectively.

In this regard, the mapped probability density at each energy state-wise is $n_N \propto \sqrt[3]{E_N} \equiv \sqrt[3]{NE_g} \propto N(eV_i)$. At a fixed boundary length, the probability of finding the particle within different potential boundary values can also be interpreted from the above relation. Fon instance, using the quantization mapping, the estimated ground state energy values are 30.868, 61.736, 92.604, 123.472 meV, …, under the fixed potential boundary values of 1, 8 ($=2^3$), 27 ($=3^3$), 64 ($=4^3$), 125 ($=5^3$) eV, …, respectively, and hence the governed relationship between the ground state energy $(E_1 = E_g)$ and voltage-driven potential boundary is described by,

$$\left(E_g\right)_1 = \frac{1}{2}\left(E_g\right)_{2^3} = \frac{1}{3}\left(E_g\right)_{3^3} = \cdots = \frac{1}{N}\left(E_g\right)_{N^3} \qquad (8)$$

Therefore, the generalized relation for the $N^{th}$ quantum state energy under different given potential values is,

$$(E_N)_1 = \frac{1}{2}(E_N)_{2^3} = \frac{1}{3}(E_N)_{3^3} = \cdots = \frac{1}{N}(E_N)_{N^3} \qquad (9)$$

On the other hand, the expected energy quantization under the voltage-driven potential boundary conditions for 2D non-Dirac (nonrelativistic dynamics) transport materials can be mapped by the relation of $E_N \rightarrow NE_g \propto N^2(eV_i)$.

## 4. QUANTIZATION MOBILITY

By applying our quantization procedures (see postulates) to the generalized Einstein relation under the voltage-driven boundary condition of $U$=$qV_g$ (refer Tables 1 and 2; i.e., $n = n_i N^3$, $E_F = \hbar v_F k_F = N\hbar v_F k_i \equiv E_{F,i}N$), we govern the quantum effect on the diffusion-mobility relation for 2D Dirac materials as,[9,13,19]

$$\frac{\mu}{D} = \frac{dn}{dE_F}\frac{1}{n}q \rightarrow \frac{3q}{E_{F,i}}\frac{1}{N} = \frac{3q}{\hbar v_F k_{F,i}}\frac{1}{N} \quad (10)$$

where, $E_F$ is the Fermi energy. It is observed that the D/μ ratio is linearly proportional to the principal quantum number (*N*). According to the diffusion coefficient, the mobility quantization can be expressed via Eqn. 10. As described in a previous study,[9] the entropically varying diffusion equation for Dirac dynamic system is $D(h_S) = D_i exp\left(\frac{2h_S}{d+3}\right)$, where $d$ is the dimension. In this study, the principal quantum number (*N*) is associated with the voltage-driven differential entropy ($h_S$) and is mapped by the relation of $N = exp\left(\frac{h_S}{5}\right)$. In such a way, the quantization effect on the diffusion coefficient (for 2D Dirac materials) for the potentially bounded systems is mapped by,

$$D(h_S) = D_i exp\left(\frac{2h_S}{5}\right) \rightarrow D_i N^2 \quad (11)$$

It is to be noted that, for bounded electronic systems, the diffusion coefficient can be expressed in terms of $N^2$ (i.e., $D(N) = D_i N^2$. The energy gap between the existing electronic levels (quantized) is increased while the potential well increases from zero to an infinite value. Accordingly, the transport parameters like diffusion and mobility are follows the above quantization rules. On the other hand, the charge diffusion in the free electron (unbounded) solids is generally equated by $D(h_S) = D_i exp\left(\frac{2h_S}{5}\right)$. Depending on the boundary conditions, the $h_S$ value will be expected, which is logarithmically aligned with the Gaussian width of the carrier wave packet, $h_S = ln\left(\sqrt{2\pi exp(1)}\ \sigma_{GW}\right)$. By comparing the above Eqns. 10 and 11, the quantum mobility is treated here as,

$$\mu_{2D,Dirac} = \left(\frac{3q}{E_{F,i}} D_i\right) N = \left(\frac{3qv_F}{2\hbar}\right) \frac{\tau_i}{k_{F,i}} N \cong \mu_i N \qquad (12)$$

Since, the diffusion coefficient is $D = \frac{v_F^2 \tau}{2}$.

**Table 3.** $N^3$ rule-based voltage-driven quantization effect on the entropy-ruled μ/D ratio in a different potentially bounded graphene system obeys the inverse quantum relation of 1/*N* (inversely proportional to the principal quantum number, *N*). The calculated diffusion coefficient and mobility follow the quantum rules of $N^2$ and *N*, respectively, under the applied gate voltage-driven boundary values of $U(N)=N^3 eV_g$. It is noted that the mobility has a linear relationship with the existing energy levels. The mobility is principally associated with the DOSP, $\frac{dh_S}{dE_N} = \frac{1}{E_{F,i}} \frac{dh_S}{dN} \approx \frac{1}{E_{F,i}} \frac{h_S}{N}$.

| Applied Gate-Voltage ($V_g$) | Energy State (*N*) | Wavevector ($10^7$ m$^{-1}$) | Einstein μ/D Ratio (V$^{-1}$) | Diffusion Constant ($10^{10}$ cm$^{-2}$s$^{-1}$) | Quantum Mobility (cm$^{-2}$V$^{-1}$s$^{-1}$) |
|---|---|---|---|---|---|
| 1 ($=1^3$) | 1 | 4.689 ($=k_1$) | $(\mu/D)_1 = 97.19/1$ | $D_1$ ($=1^2 \times D_1$) | $\mu_1$ ($=1\times \mu_1$) |
| 8 ($=2^3$) | 2 | 9.379 ($=2k_1$) | $(\mu/D)_2 = 97.19/2$ | $D_2$ ($=2^2 \times D_1$) | $\mu_2$ ($=2\times \mu_1$) |
| 27 ($=3^3$) | 3 | 14.693 ($=3k_1$) | $(\mu/D)_3 = 97.19/3$ | $D_3$ ($=3^2 \times D_1$) | $\mu_3$ ($=3\times \mu_1$) |
| 64 ($=4^3$) | 4 | 18.759 ($=4k_1$) | $(\mu/D)_4 = 97.19/4$ | $D_4$ ($=4^2 \times D_1$) | $\mu_4$ ($=4\times \mu_1$) |
| 125 ($=5^3$) | 5 | 23.463 ($=5k_1$) | $(\mu/D)_5 = 97.19/5$ | $D_5$ ($=5^2 \times D_1$) | $\mu_5$ ($=5\times \mu_1$) |
| 216 ($=6^3$) | 6 | 28.134 ($=6k_1$) | $(\mu/D)_6 = 97.19/6$ | $D_6$ ($=6^2 \times D_1$) | $\mu_6$ ($=6\times \mu_1$) |

The governed mobility (Eqn. 12) is suitable for potentially bounded 2D Dirac systems/devices. For open (unbounded) systems, the general version of mobility can be obtained by

$$\mu_{Dirac,d} = q\left(\frac{d+1}{d+3}\right) \frac{dh_S}{dE_F} D(h_S) \rightarrow q\left(\frac{d+1}{d+3}\right) \frac{dh_S}{dE_F} \frac{v_F^2 \tau_{h_S}}{2} \qquad (13)$$

Therefore, the mobility of 2D Dirac transport systems is $\frac{3q}{5} \frac{dh_S}{dE_F} D(h_S)$. That is, the mobility is directly proportional to the diffusion constant times of the DOSP. It has been observed that the voltage-driven diffusion enhancement is larger than that of mobility. For instance, the required voltage (drift) to double the electron mobility is $2^3$ volts. In this case, the diffusion is 4-times increased from the initial value (see Table 3). The results suggest that the applied bias voltage significantly improves the charge distribution in length scale (charge flux over the space), which leads to diffusion dominant over the charge mobility in such quantum systems. Therefore, the mobility-diffusion ratio (μ/D) falls in an inverse proportional relation to the principal quantum number, 1/*N* (see Eqn. 10). That is, the field-response effect on the diffusion coefficient is larger than the mobility, which is in agreement with the earlier studies (see cross references in ref. 13).[14,25-27] Under quantization mapping ($U=N^3qV_i$), the diffusion coefficient and mobility are increased by orders of $N^2$ and $N$, respectively. Interestingly, the quantum-to-fractional quantum behaviour shift is observed at degenerate conditions ($T\rightarrow0$; $E_F>>k_BT$) and boundary conditions, which can be analysed by differential entropy-associated wavevector propagation. According to the boundary value, the shift is noted from the Gaussian width of the carrier wave packet, which will influence the wave dynamics via differential entropy.

## 5. CONCLUSIONS

From our investigations through the proposed postulates, it is concluded that the bounded and unbounded massless Fermion dynamics in monolayer graphene have been incorporated using differential entropy ($h_S$)-ruled wavevector ($k$) mechanics and the Klein paradox. The formation of electron-hole puddles due to disorder and thermodynamic effects in nearer to the charge neutrality point is addressed by the entropically varying charge density equation. The populated electronic states at different potential boundaries are hereby mapped with the DOSP, which is principally defined as the energy (in terms of wavevector, for Dirac dynamics) scaled differential entropy of a given electronic system/devices. The validity and breakdown of the Klein paradox is explained for a free particle and a system of bounded Fermionic particles through an entropy-ruled wavevector mechanism. The introduced "$h_S$-ruled $k$" and "$h_S$-ruled $N$" descriptions, such as $k(h_S) = k_i exp\left(\frac{h_S}{5}\right)$ and $N = exp\left(\frac{h_S}{5}\right)$ are here implemented to generalize the electron dynamics in both unbounded and potentially bounded Dirac systems. By quantization procedure, the existence of energy levels with respect to the potential well is mapped and it obeys the relation of $E_N = NE_g \propto N^3(eV_i) \leftrightarrow U(V_g)$; where, $E_g$, $N$, and $U(V_g)$ are the ground state energy, principal quantum number ($N = 1, 2, 3, \ldots$), and the gate-voltage-driven potential well energy, respectively. The probability for finding the particle inside the boundary (in terms of charge density) at different energy states can be inferred by the mapping equation $P_{E_N}: n_N \propto \sqrt[3]{(E_N)_{U=eV}} \propto N(eV_i)$. From our analysis, the main observation is that for any quantum systems, the excitonic property and relevant carrier energy transition (from lower ground state to higher/excited state) can be triggered by the applying voltage-driven potential using the mapped ratio-wise relation of $N_{Q.St.} \leftrightarrow N^3_{U=eV_g}$; where, $N = 1, 2, 3, \ldots$, and $V_g$ is the gate-voltage, respectively. Accordingly, the calculated electron density, diffusion coefficient, and mobility in the potentially bounded Dirac quantum systems are increased by orders of $N^3$, $N^2$ and $N$, respectively. The revealed quantization mapping via entropy-ruled wavevector mechanics for the Dirac system hopefully will initiate a new dimensional fine-way approach to design the ultrafast electron devices for various potential applications like electronic, optical, energy, thermoelectric, and excitonic properties, including electron transfer kinetics in quantum-confined semiconductors, etc.

## NOTES

The author declares no competing financial interest.

## ACKNOWLEDGMENT

The author KN is grateful to the authors of references 6-9 and 13, which help design the postulates and validate the proposed entropy-ruled wavevector mechanism for the Dirac transport systems.